\documentstyle[11pt,paspconf]{article}

\begin{document}

\title{Observational Effects of Strange Stars}
\author{T. Lu}
\affil{Astronomy Department, Nanjing University, Nanjing 210093, China}

\begin{abstract}
In this talk, after briefly reviewing some historical remarks 
concerning strange stars, the achievements in physics and dynamical 
behavior of strange stars are discussed. Especially, various observational 
effects in distinguishing strange stars from neutron stars such as 
mechanical effects, cooling effects, phase transition and 
related interesting phenomena are stressed.
\end{abstract}


\section{Introduction}

Though some ideas of quark stars were proposed as early as in 1969,
5 years after Gell-Mann's prediction of the particle ``quark'', and 
Bodmer even touched the concept of strange (quark) matter in 1971, no 
great achievements had been made until 1984.

In 1984, Witten, and somewhat later 
Farhi \& Jaffe, predicted the existence of strange 
stars by proving their energies lower than neutron stars for rather wide
range of QCD parameters in the MIT bag model, hence, the 
stable one might be strange star, not neutron star. The conjecture
has attracted a lot of attentions, which even lead Alcock et al. (1986)
to the argument that there may be no neutron stars, only strange stars!
Do all known neutron stars or only some or even none of them be really 
strange stars? The final solution should be given by observation. So, the 
most important and difficult task is to study carefully the 
observational effects to distinguish strange stars from neutron stars. 

Also in 1984, Wang \& Lu pointed out the extremely high efficiency of the 
non-leptonic weak process ($u d \longleftrightarrow u s$) within
quark (strange) matter in damping away the radial vibration of a neutron 
star with strange quark core or of a strange star, 
a key point to the dynamics of strange matter. This damping implies the very 
high bulk viscosity in strange matter (Sawyer 1989; Haensel 1991; 
Madsen 1992). This effect
is closely related with observations.

In 1986, Alcock et al. studied the structures of the strange stars. 
They suggested the existence of a crust over the surface of a strange 
star similar to the outer crust of a conventional neutron star. As 
the radiation is mainly determined by the magnetosphere, which is
closely related with the stellar crust, almost identical outer crust of
strange star and neutron star makes it difficult to distinguish 
between them. If strange star really exists, one should find out
the effects of the differences in the stellar interior, such as 
rotation, cooling, phase transition and else, on the radiations.
\newpage

It may be very interesting to mention that the wrong discovery of a
0.5 ms optical pulsar within SN1987A in 1989 promoted the progressive 
development in the field of strange stars, as the sub-milli-second 
period for a neutron star is difficult to achieve based on reasonable 
equation of state for normal neutron matter (Frieman 1989; Glendenning 1989),
while strange stars can reach much shorter period than neutron stars.

In 1991, a successful international workshop on Strange Stars was 
held in Denmark. Details are referred to the Proceedings of that 
Workshop on Strange Quark Matter in Physics and 
Astrophysics (ed. Madsen \& Haensel) ---- Nucl. Phys. 
(Proc. Suppl.), 1991, 24B. Since then, strange star has been one of
standard topics in many meetings on physics and astrophysics.
And a lot of achievements in this field have been obtained, see, for
example, Dai and Lu (1994b).

\section{Structure of Strange Stars in Comparison with Neutron Stars}
 
Alcock et al.(1986a) were the first
to give a rough description of the structure of a strange star near its 
surface based on simple Thomas-Fermi approximation. Due to $m_{\rm s} \neq 
m_{\rm u,d}$, the net quark charge in the strange star is slightly positive and
must be balanced by electrons. As quarks bound through strong interaction,
the strange matter should have very sharp surface with thickness of the 
order of $1$ fm. On the contrary, the electrons, bound by the Coulomb force,
can extend several hundred fermis beyond the quark surface. So, a strong 
electric field, about $\sim 10^{17}$ V cm$^{-1}$ and outwardly directed,
will be established in a thin layer of several hundred 
fermis thick above the strange matter surface. Normal nuclear matter 
will be prevented by this strong field from falling into the strange 
matter and be supported as a crust. This argument leads to the structure
of a strange star as composed of two parts, a big strange core and 
a thin nuclear crust, with a gap in between.

As a rough estimation, consider the simple degenerate case (low 
temperature), the number density of electrons 
$n_{\rm e}$ can then be locally related with the electron Fermi 
momentum $p_{\rm e}$
\begin{equation}
n_{\rm e} = p_{\rm e}^{3}/3\pi^{2}
\end{equation}
(in the unit system of $c = \hbar = 1$).
Equilibrium implies that $p_{\rm e} = V$, here $V$ is the electrostatic 
potential energy. Then the local charge distribution generates the 
potential, so Poisson's equation reads (Alcock et al. 1986):
\begin{equation}
\frac{d^{2} V}{d z^{2}}=\left\{
     \begin{array}{llr}
       {{4 \alpha}/{3 \pi}}(V^{3}-V_{q}^{3}), & z \leq 0,  \\ \\
       {{4 \alpha}/{3 \pi}}V^{3}, & 0 < z \leq z_{\rm G}, \\ \\
       {{4 \alpha}/{3 \pi}}(V^{3}-V_{\rm c}^{3}), &  z_{\rm G} < z, 
     \label{eq:alcock}
     \end{array}
  \right.
\end{equation}
where $z$ is a space coordinate measuring the height above the quark 
surface, $\alpha$ the fine-structure constant, 
$V_{\rm q}^{3}/3\pi^{2}$ the quark charge density inside the quark matter,
$V$ the potential energy, and $V_{\rm c}$ the electron Fermi momentum 
deep in the crust, it describes the positive charge 
density of the ions within the crust, $z_{\rm G}$ the gap width between the 
strange matter surface and the base of the crust. In fact, $V_{\rm q}$ 
and $V_{\rm c}$ are the boundary values of the above equation:
$V \rightarrow V_{\rm q}$ as $z \rightarrow - \infty$, $V \rightarrow
V_{\rm c}$ as $z \rightarrow + \infty$. Meaningful solutions
to the above equations occur only if $V_{\rm c} < V_{\rm q}$.

As the crust getting bigger and bigger, the density at the base of the
crust ($\rho_{\rm crust}$) will be higher and higher. When 
$\rho_{\rm crust}$ reaches the neutron drip point (that is, 
$\rho_{\rm crust} = \rho_{\rm drip} \sim 4.3 \times 10^{11}$ g cm$^{-3}$,
$A=118$, $Z=36$), neutrons will begin to drip out. Being electrically
neutral, they will fall freely 
into the strange core and be deconfined to be strange matter. Hence,
$\rho_{\rm drip}$ sets the absolute limit for the density at the base of
the crust. As is well known, neutron star has inner crust with density
from $\rho_{\rm drip}$ to nuclear density $\rho_{\rm nucl}$, this  
range covers densities higher than $\rho_{\rm drip}$. So, strange stars can
only have outer crust, not inner crust. 

However, though the strange star 
can be bound even without gravitation, its crust
can not be bound to the strange core without gravitation. 
Could the density at the base of the crust reach $\rho_{\rm drip}$, or 
could it only reach some value $\rho_{\rm crust} \ll \rho_{\rm drip}$?
This question may be solved by taking mechanical balance between
electrical force and gravitation on the crust into consideration.
In fact, there are two kinds of forces acting on the crust: one is the 
electric force by the strong electric field acting on the charge of
the crust, the other is the gravitational force acting on its mass.
The former favors the crust, while the latter disfavors it. Alcock
et al. (1986) simply argued that the force of strong electric field on a 
single nucleus overwhelms the force of gravity which then could be neglected. 
This led them to the conclusion that the only limit on the crust density
is determined by the neutron drip. Based on this point of view, 
Glendenning et al. (1992, 1995a and 1995b) studied the series of
core-crust systems. This series is very interesting which covers the 
whole series of strange counterparts of neutron stars and white dwarfs.
Massive strange core (of mass $\sim M_{\odot}$) with thin crust is the
counterpart of neutron star, this is so called strange star. Less
massive core with very, very thick crust is the counterpart of white 
dwarf, this is called as strange dwarf.

Note, the crust (taking electrons into account) is only 
weakly charged (nearly neutral), the electric force on the crust is
rather weak, mechanical balance should be held between electric and 
gravitational forces on the crust, not only on a single nucleus (Huang 
and Lu, 1997a, 1997b). Based on 
this analysis, we recalculate the question in a consistent way, and 
prove that the gap height will obviously decrease as  
the crust density increases and the crust can not be supported further 
for a density $\rho_{\rm crust}$ at the base of the crust far lower 
than the neutron drip density, quite different from results previously 
known. Calculations indicate that at $\rho_{\rm crust} = 8.3 \times
10^{9}, 3.3 \times 10^{10}, 2.1 \times 10^{11}$ and $3.3 ^{11}$ g$/$cm$^{3}$,
we have $z_{\rm G} = 760, 380, 63$ and $7$ fm respectively.
Thus, $z_{\rm G}$ is already very small when $\rho_{\rm crust}$ is
still far lower than $\rho_{\rm drip}$. In fact, when $\rho_{\rm crust}$
equals $\rho_{\rm drip}$, a meaningful $z_{\rm G}$ (positive) can not
be found. The maximum density at the base of the crust is restricted by 
the electric field rather than the neutron drip point.

As the calculation of Alcock et al. (1986) was based on 
Eq.(\ref{eq:alcock}) which regard the crust as a continuum, while
the lattice spacing in the crust is $\sim 200$ fm, if 
$z_{\rm G} \le 200$ fm, the above smooth description of crust would not
be correct. So, as Alcock et al. and many other authors adopted, we will 
also use $200$ fm as the limit of the crust height, the corresponding
density at the base of the crust would be
\begin{equation}
\rho_{\rm crust} = 8.3 \times 10^{10} {\rm g cm}^{-3} \sim \rho_{\rm drip}/5
\ll \rho_{\rm drip}.
\end{equation}
This means that at a crust density still far lower than the neutron 
drip density, the crust would begin to break down. If we take
$z_{\rm G} = 200$ fm as the gap limit, the maximum mass of a crust
should be about 
\begin{equation}
M_{\rm crust} \sim 3.4 \times 10^{-6} {\rm M}_{\odot}. 
\end{equation}
For the most interesting stellar mass range
of 0.01 to 2 $M_{\odot}$, 
the calculated crust mass would be kept to be the value of Eq.(4), one
order of magnitude lower than previously known (based on $\rho_{\rm crust}
= \rho_{\rm drip}$).

\section{Observational Effects of Strange Stars}
 
Here, we will concentrate our discussions on observational effects
of strange stars in order to distinguish them from neutron stars and
will not discuss strange dwarfs.

\subsection{Global Properties} 

The equation of state of strange matter is different from that of neutron 
matter. The global properties of strange star then should also be different 
from that of neutron star. These differences lead to different dependence
of stellar mass on stellar radius for strange star and neutron star.
For neutron star, stellar mass decreases with 
increasing radius, while for strange star, mass increases with increasing
radius.
So big differences should seem to be observed 
``easily''! Unfortunately, for typical masses $M \sim 1.4 M_{\odot}$,
both kinds of stars will give almost the same results! Thus, it is 
still difficult to distinguish strange star and neutron star based on
these global properties.

\subsection{Radial Vibrations}

Wang and Lu (1984) was the first to study the damping effects of the
vibrations in the strange core of a neutron star or in a strange star.
For radial vibrations of a star, the mean dissipation rate of the
vibrational energy per unit mass can be expressed as
\begin{equation}
(\frac{dw}{dt})_{\rm ave} = -\frac{1}{\tau}\int_{\rm period} Pdv 
\label{eq:dwdt}
\end{equation}
here $v = v_{0} + \triangle v \cos (2\pi t/ \tau)$ denotes the volume
per unit mass, and $P$ is the pressure. Note, $P$ is closely related
with particle number density. During radial vibration, the volume per
unit mass will change. Even if at a moment the strange matter is in
equilibrium, weak processes $u s \longleftrightarrow u d$ will
make it deviate from equilibrium during volume variation. Wang and 
Lu (1984, 1985) pointed out that in the case of strange matter the 
dissipation rate (Eq.(\ref{eq:dwdt})) is very large, several orders 
of magnitude larger than the case of neutron matter. This dissipation
will very effectively damp away the radial vibrations of strange stars
or even neutron stars with strange core. The time scale of the damping
is in the order of milli-seconds or tens of milli-seconds. This is a
very sensitive effect for the strange matter. If you could
observe this short time scale radial vibration damping, you could say
that there would be strange matter within the star.

\subsection{Bulk Viscosity}

The strong damping effect of the radial vibration of strange star 
implies the high bulk viscosity in strange matter. Haensel, Zdunik 
and Schaeffer (1989) related this damping with the viscosity 
(they defined the second viscosity). Sawyer (1989) and Madsen (1992)
used this damping effect further to define a coefficient of bulk viscosity, 
$\zeta$, through the following equation
\begin{equation}
(1/v_{0})(dw/dt)_{\rm ave} = - (1/2) \zeta (2 \pi / \tau)^{2} 
(\triangle v / v_{0})^{2}, 
\label{eq:zeta}
\end{equation}
This coefficient provides a more convenient way to study the dynamical
behavior of strange matter and strange star. The influence of temperature,
density, frequency and s-quark mass on the damping effect could be
expressed as the coefficient to be a function of these physical 
quantities. These dynamical behaviors have been studied further by 
Heiselberg (1992), Benvenuto, Horvath (1991), Horvath, Vucetich, 
Benvenuto (1993), Goyal et al. (1994), Dai and Lu (1995, 1996).

\subsection{Spin Rate}

Spin rate or rotation period of a neutron star or a strange star can 
be measured very accurately. 
The different relations between mass and radius seems to make strange 
star and neutron star to have different 
moment of inertia, so to have different spin hehavior. Unfortunately, 
for the typical mass 1.4 M$_{\odot}$, the radius of a strange star is 
approximately the same as of a neutron star. 

However, strange matter and neutron matter  
have different viscous properties (Wang \& Lu 1984, 1985; Sawyer 1989). 
This may make important influence to the dynamical behavior of these stars.
As is well known, the spin rate of a pulsar should be limited by
Kepler condition. When its spin rate exceeds the Kepler speed 
(Lattimer et al. 1990), the pulsar will be unstable, matter will flow
out from its equator. However, gravitational radiation reaction
instabilities is supposed to set the spin rate limit for neutron stars
far lower than the Kepler limit. The shear (Haensel \& Jerzak 1989) and 
especially the bulk (Wang \& Lu 1984, 1985; Sawyer 1989) viscosities can
damp away the gravitational radiation reaction instabilities. Due to
very high viscosity in strange matter, the spin rate of strange stars
can be much closer to the Kepler limit than that of neutron stars
(Zdunik 1991; Madsen 1992). This provides an important way to find
strange stars observationally. If once a pulsar with sub-milli-second 
period could be discovered, it may probably be a strange star rather
than a neutron star.

\subsection{Cooling}

Strange star has its neutrino emissivity stronger than neutron star,
it will cool quick. Pizzochero (1991), Dai and Lu (1994a) pointed out, 
the surface temperature of strange stars from age of 1 to $10^{5}$
years is usually lower than that of neutron stars. Their surface
temperature could be measured through X-ray data analyses. This 
temperature difference provides an important method to distinguish 
strange stars from neutron stars observationally. From X-ray data 
analyses, the observed temperatures of some pulsars, such as Crab and 
Vela pulsars, seem to be in agreement with that predicted from 
neutron star cooling. However, recent data from ROSAT show that 
the surface temperature of PSR 0656+14 seems far lower than that
predicted from neutron star cooling, but closer to that predicted
from strange star cooling. Perhaps, this may imply that PSR0656+14
is a strange star? (Dai, Lu, Song, Wei 1993)

Recently, Schaab et al. (1997) studied the 
neutrino emissivity and pointed out that the direct Urca process can 
be forbidden not only in neutron stars, but also in strange stars.
In this case, strange stars may cool slowly and their surface 
temperatures may be more or less indistinguishable from neutron stars.
Based on the suggestion of Bailin and Love (1979, 1984), quarks
in stars may eventually form Cooper pairs, thus superfluid may also
exist in strange stars. This would suppress the neutrino emissivities,
slowdown the cooling and provide further similarities between strange 
stars and neutron stars. Schaab et al. (1997) calculated this case in
some detail. They found that within the first $\sim$ 30 years 
after birth, strange stars may cool more rapidly than neutron stars. 
If this picture is correct, people should try to find strange star 
within the SN1987A. 

\subsection{Phase Transitions}

How can a neutron star be converted into a strange star?
If there happens to be a strange matter seed at some place within a neutron
star, neutrons, being charge neutral, can easily diffuse into the 
seed and be deconfined to be strange matter. Here, the seed may come 
from outside, or more possibly be formed at a locally high density 
place within a neutron star. The strange 
matter seed can get bigger and bigger until the whole star 
being converted into a strange star (Olinto 1987). Olesen and Madsen (1991)
studied the processes of conversion of neutron matter to strange
matter based on the diffusion mechanism of Olinto (1987). They considered
different vacuum energy densities $B$ and different masses of neutron stars
and proved the time scale of conversion in the range of 0.1 s to several
mimutes.

Dai, Wu and Lu (1995, 1996) studied an example of the conversion
from neutron star to strange star based on general relativity. Consider a
hybrid star with baryon mass of 1.4 $M_{\odot}$, its interior is a 
strange matter core of radius $r_{\rm s}$, and its exterior is 
neutron matter. By integrating the Tolman-Oppenheimer-Volkoff
equation, the radius $R$, the moment of inertia $I$ and the gravitational 
mass $M$ of the hybrid star are calculated as a function of $r_{\rm s}$.
Here, we chose BPS (Baym et al. 1971a,b), V$_{\gamma}$ (Tsuruta \& 
Cameron 1966) and Wb (Waldauser et al. 1987) as soft, medium stiff and 
stiff equations of state for neutron matter; and chose typical 
values of $m_{\rm s} = 200$ MeV, $B = 57$ MeV fm$^{-3}$, and 
two values of $\alpha_{\rm c} = 0$ and $0.6$ as two kinds of equations of 
state in MIT bag model for strange matter.  
As neutrons diffusing into the strange core (Olinto 1987), the 
core will get bigger and bigger, $r_{\rm s}$ will increase with time
until the whole star becoming a strange star ($r_{\rm s} \longrightarrow
R$, R is the stellar radius). This may appear as a
giant glitch observationally. Note that
the amplitude of the glitch is mainly determined by the state of
neutron matter, while the time scale of the glitch is mainly 
determined by the state of strange matter. This provides a way to
observe strange stars and study the equations of state of the stars.

There might be hydrodynamical instability in the above diffusion based
slow process (Horvath, Benvenuto, 1988). The conversion of neutron
matter to strange matter might also proceed in a detonation-like rapid 
mode. In such a mode, there will appear two stages, namely, the 
formation of two-flavor quark matter and the weak decay processes
leading to the formation of strange matter. The time scale for the
conversion of a neutron star to a strange star in this rapid mode is 
usually smaller than 1 s.

\subsection{Neutrino Bursts}

The phase transition of two-flavor quark matter to three-flavor 
(strange) quark matter was studied based on the assumptions that
neutrinos escape freely out of the star as soon as they are produced and
the temperature in the interior is constant during the conversion
(Dai, Lu \& Peng 1993a,b).  They proved that the time scale of the
transition is less than 1 $\mu$s. A strong neutrino burst will 
happen during the transition through semi-leptonic weak processes 
participated by s quark. 
The neutrino burst 
duration is very short, less than 1 $\mu$s. If a phase transition of 
neutron matter to strange quark matter happens in the interior of 
a neutron star of mass $M = 1.4 M_{\odot}$, there will be a neutrino
burst with energy of several $10^{51}$ ergs emitted.

\subsection{Supernova Explosions}

Gentile et al. (1993) studied the phase transition of nuclear matter into
strange matter in a supernova core. As suggested by Alcock et al. (1986), 
the process is very likely to consist of two 
distinguishable processes: nuclear matter to two-flavor quark matter and
two-flavor quark matter to strange matter.

Dai, Peng \& Lu (1994, 1995) studied further the phase transition in the 
supernova core. Note, this transition is quite different from that in 
a neutron star. Neutrinos can be emitted freely in a neutron star, and
will be trapped in a supernova core, as their mean free path being 
much smaller than the radius of a supernova core. Once neutrino trapping 
occurs, the collapse will be an adiabatic rather than isothermal process.
The typical temperature in a supernova core should be about $10^{11}$ K,
much, much higher than in the case of a neutron star. They found that
the time scale of the transition is below $10^{-7}$ s. Due to this 
conversion, both the temperature in the inner core and the neutrino 
energy in the whole core increase significantly. These results could
enhance both the probability of success for a supernova explosion and
the total energy of the revived shock wave, and could also affect the
cooling of a newborn neutron star. 

Recently, Anand et al. (1997) 
re-studied this transition in a detail. They have systematically taken
into account the effect of strong interactions perturbatively to order
$\alpha_{c}$ and the effect of finite temperature and strange quark mass.

\subsection{Gravitational Radiation}

Chau (1967) pointed out that an oscillating and rotating compact 
object can radiate strong gravitational waves. As an oscillating and 
rapidly rotating strange star could be produced after a 1.4 solar 
masses neutron star accreting 0.5 solar masses from its companion 
(Cheng and Dai 1996), this strange star could thus radiate strong
gravitational radiation. Cheng and Dai (1997) pointed out that the
time scale of the damping due to gravitational radiation is in the 
order of several $10^{-2}$ s. Comparing with other mechanisms of
damping, such as due to viscosity, neutrino emission and others,
they indicated the importance of gravitational radiation.

\subsection{$\gamma$-ray Bursts}

The concept of strange star may provide a new possibility
to explain the mysterious phenomena of $\gamma$-ray bursts.
In fact, the original motivation of the early research on
damping of radial oscillation of strange star (Wang \& Lu, 1984, 1985)
is to try to explain the $\gamma$-ray bursts.

The observations by CGRO(BATSE) and also the recent discovery of afterglow
by BeppoSAX (Paradijs et al. 1997) all indicate that the $\gamma$-ray 
bursts are at cosmological distances. Inevitably, there should be 
fireballs as their sources which could be produced, for example, by 
the neutron star
mergers. The heavy baryon contamination in such a merger will be a very 
serious difficulty to explain sufficient $\gamma$-ray intensity.
Cheng and Dai (1996) argued that the conversion of neutron stars to 
strange stars could be possible origin of $\gamma$-ray bursts.
There are baryons only in the very thin crust of a strange star,
in such a model, baryon contamination will not be a problem. This 
mechanism to explain $\gamma$-ray bursts may be attractive.

\section{Discussions}

We have discussed a lot of observational effects to distinguish
strange stars from neutron stars. As the theory on strange 
matter is still very rough, these effects can not yet be regarded
as definite. However, the strange matter is as fundamental as black
hole. Though its existence has not yet been solved, further studies
will still be very important. Who will be the first to prove or 
disprove the existence of strange matter? Physicists are enthusiastic
to observe strangelet in the accelerator laboratory and astrophysicists
are equally enthusiastic to find strange stars by astronomical methods.

\acknowledgments

This work 
was supported by the National Natural Science Foundation and the 
Foundation of the National Committee for Education of China.


\begin{references}
\reference Alcock, C., Farhi, E., \& Olinto, A. 1986, \apj, 310, 261
\reference Anand, J.D., Goyal, A., Gupta, V.K., Singh, S. 1997, \apj,
    481, 954
\reference Bailin, D., \& Love, A. 1979, J. Phys. A, 12, L283
\reference Bailin, D., \& Love, A. 1984, Phys. Rep., 107, 325
\reference Baym, G., Bethe, H.A., \& Pethick, J.C. 1971a, Nucl. Phys., A175, 225
\reference Baym, G., Pethick, J.C., \& Sutherland, P. 1971b, \apj, 170, 299
\reference Benvenuto, O.G., \& Horvath, J.E. 1991, \mnras, 250, 679
\reference Bodmer, A.R. 1971, Phys. Rev., 4, 1601
\reference Chau, W.Y., 1967, \apj, 147, 667
\reference Cheng, K.S., \& Dai, Z.G. 1996, Phys. Rev. Lett., 77, 1210
\reference Cheng, K.S., \& Dai, Z.G., 1997, \apj, to be published
\reference Dai, Z.G., \& Lu, T. 1995, Acta Astronomica Sinica, 36, 165
\reference Dai, Z.G., \& Lu, T. 1996, Zeit f. Phys. A, 355, 415
\reference Dai, Z.G., Lu, T., \& Peng, Q.H. 1993a, Phys. Lett., 319B, 199
\reference Dai, Z.G., Lu, T., \& Peng, Q.H. 1993b, Acta Physica Sinica, 42, 1210
\reference Dai, Z.G., \& Lu, T. 1994a, Acta Physica Sinica, 43, 198
\reference Dai, Z.G., \& Lu, T. 1994b, Progress in Physics, 14, 327
\reference Dai, Z.G., Lu, T., Song, L.M., \& Wei, D.M. 
             1993, Acta Astronomica Sinica, 34, 225
\reference Dai, Z.G., Peng, Q.H., \& Lu, T. 
             1994, Acta Astronomica Sinica, 35, 337
\reference Dai, Z.G., Peng, Q.H., \& Lu, T. 1995, \apj, 440, 815
\reference Dai, Z.G., Wu, X.J., \& Lu, T. 1995, Ap\&SS, 232, 131
\reference Dai, Z.G., Wu, X.J., \& Lu, T. 1996, Acta Astronomica Sinica, 37, 21
\reference Farhi, E., \& Jaffe, R.L. 1984, Phys. Rev. D, 30, 2379
\reference Frieman, J.A., \& Olinto, A.V. 1989, Nature, 341, 633
\reference Gentile, N.A., et al. 1993, \apj, 414, 701
\reference Glendenning, N.K. 1989, Phys. Rev. Lett., 63, 2629
\reference Glendenning, N.K., \& Weber, F. 1992, \apj, 400, 647
\reference Glendenning, N.K., Kettner, Ch., \& Weber, F. 1995a, Phys. Rev. Lett.,
                74, 3519
\reference Glendenning, N.K., Kettner, Ch., \& Weber, F. 1995b, \apj, 450, 253
\reference Goyal, A., Gupta, V.K., Pragya, \& Anand, J.D. 1994, Zeit. f. Phys. A,
                349, 93
\reference Haensel, P. 1991, Nucl. Phys. (Proc. Suppl.), 24B, 23 
\reference Haensel, P., \& Jerzak, A.J. 1989, Acta Phys. Pol. B 20, 141
\reference Haensel, P., Zdunik, J.L., \& Schaeffer, R. 1989, A\&A, 217, 137
\reference Heiselberg, H. 1992, Physica Scripta, 46, 485
\reference Horvath, J.E., Benvenuto, O.G. 1988, Phys. Lett. B, 213, 516
\reference Horvath, J.E., Vucetich, H., \& Benvenuto, O.G. 
                 1993, \mnras, 262, 506
\reference Huang, Y.F., \& Lu, T. 1997a, Chin. Phys. Lett., 14, 314
\reference Huang, Y.F., \& Lu, T. 1997b, A\&A, 325, 189
\reference Lattimer, J.M., Prakash, M., Masak, D., \& Yahil, A.
                1990, \apj, 355, 241
\reference Lu, T. 1991, Quark-Gluon Structure of Hadrons and Nuclei, ed.
       L.S. Kisslinger and Qiu Xijun, International Academic 
       Publishers, 248
\reference Madsen, J. 1992, Phys. Rev. D, 46, 3290
\reference Olesen, M.L., Madsen, J. 1991, Nucl. Phys. B (Proc. Suppl.), 24B, 170
\reference Olinto, A.V. 1987, Phys. Lett., 192B, 71
\reference Paradijs, J.van, et al. 1997, Nature, 386, 686
\reference Pizzochero, P.M. 1991, Phys. Rev. Lett., 66, 2425
\reference Sawyer, R.S. 1989, Phys. Lett., 233B, 412
\reference Schaab, C., et al. 1997, \apj, 480, L111
\reference Tsuruta, S., \& Cameron, A. 1966, Can. J. Phys., 44, 1895
\reference Waldauser, B., et al. 1987, Phys. Rev. C36, 1019
\reference Wang, Q.D., \& Lu, T. 1984, Phys. Lett., 148B, 211
\reference Wang, Q.D., \& Lu, T. 1985, Acta Astrophysica Sinica, 5, 59
\reference Witten, E. 1984, Phys. Rev. D, 30, 272
\reference Zdunik, J.L. 1991, Nucl. Phys. (Proc. Suppl.), 24B, 119
\end{references}
\end{document}